# Your Neighbors Are My Spies: Location and other Privacy Concerns in GLBT-focused Location-based Dating Applications

Nguyen Phong HOANG, Yasuhito ASANO, Masatoshi YOSHIKAWA

*Department of Social Informatics, Graduate School of Informatics, Kyoto University*
*Yoshida-Honmachi, Sakyo-ku, Kyoto 606-8501, Japan*
hoang.nguyenphong.jp@ieee.org, asano@i.kyoto-u.ac.jp, yoshikawa@i.kyoto-u.ac.jp

*Abstract*—Trilateration is one of the well-known threat models to the user's location privacy in location-based apps; especially those contain highly sensitive information such as dating apps. The threat model mainly bases on the publicly shown distance from a targeted victim to the adversary to pinpoint the victim's location. As a countermeasure, most of location-based apps have already implemented the "hide distance" function, or added noise to the publicly shown distance in order to protect their user's location privacy. The effectiveness of such approaches however is still questionable. Therefore, in this paper, we investigate how the popular location-based dating apps are currently protecting their user's privacy by testing three popular GLBT-focused apps: Grindr, Jack'd, and Hornet. We found that Jack'd has the most privacy issues among the three apps. As one of our findings, we also show how the adversary can still figure out the location of a targeted victim even when the "show distance" function is disabled in Grindr. Without using sophisticated hacking techniques, our proposed model (*called colluding-trilateration*) is still very effective and efficient at locating the targeted victim, and of course in a so-called "legal" manner, because we only utilize the information that can be obtained just as same as any other ordinary user. In case of Hornet, although it has adopted location obfuscation in its system, we were not only able to discover its noise-adding pattern by conducting empirical analysis, but also able to apply the colluding trilateration used in Grindr to locate the targeted victim regardless of the location obfuscation. Our study thus raises an urgent alarm to the users of those location-based apps in general and GLBT-focused dating apps in particular about their privacy. Finally, the paper concludes by suggesting some possible solutions from the viewpoints of both the LBS provider and the user considering the implementation cost and the trade-off of utility.

Manuscript received on March 11th, 2016. This work was supported by JSPS KAKENHI Grant Number 15K00423 and the Kayamori Foundation of Informational Science Advancement. The paper is a follow-up of [21], which was presented at the 18th IEEE International Conference on Advanced Communication Technology, and received the Outstanding Paper Award.

Nguyen Phong HOANG is currently a graduate student at the Department of Social Informatics, Graduate School of Informatics, Kyoto University, Japan. (corresponding author: +81-75-753-5375, fax: +81-75-753-5375, e-mail: hoang.nguyenphong.jp@ieee.org)

Yasuhito ASANO is an associate professor at the Graduate School of Informatics, Kyoto University (e-mail: asano@i.kyoto-u.ac.jp).

Masatoshi YOSHIKAWA is a professor at the Graduate School of Informatics, Kyoto University (e-mail: yoshikawa@i.kyoto-u.ac.jp).

## I. INTRODUCTION

NOWADAYS, thanks to the advancement of the Global Positioning System (GPS), most of the smart phones have a built-in GPS receiver, which assists to estimate the location information with accuracy up to just a few meters. Taking this advantage, location-based applications (*aka: location-based services or LBS*) are getting more dominant in the smart phone application market. Just a decade ago, one still had to use paper map or ask for direction when going to an unfamiliar area; while young people were surfing around online chat rooms to look for friends at that time. However, the introduction of LBS has changed our lifestyle and the way that people interact with each other thanks to its undeniable convenience. For instance, one can easily find the nearest restaurant, convenience store or shopping mall by using application like Google Map; or hang out with friends by using application like Find My Friends, etc.

### A. Privacy in General

Nevertheless, in the era of Information and Communications Technology, along with censorship and massive surveillance in cyberspace, the problem of information leakage has also become more and more severe. Tim Cook, the CEO of Apple Inc., used to say at the White House Cyber Security Summit in early 2015 that: "Privacy is a matter of life and death" [1]. As people increasingly keep more sensitive personal information in their phone, big agencies and companies like Apple have been working hard to provide the best protection to their customer's private information. However, an absolute privacy and a completely perfect countermeasure to prevent future data breaches still remain as headache matters. According to the Tenth Annual Cost of Data Breach Study published by IBM in 2015, the average consolidated total cost of a data breach is $3.8 million, increasing 23% since 2013. The report also points out that the cost incurred for each lost or stolen record that has sensitive and confidential information increased 6% from a consolidated average of $145 to $154 [2].

Among personally identifiable information, location is considered as one of the most essential factors since the leak of location information can consequently lead to the disclosure of other sensitive private information such as occupations, hobbies, daily routines, and social relationships [3]. In spite of many attack techniques [4], [5] that have been studied by the research community since then, the protection of location privacy from both LBS provider and user has not been sufficiently and appropriately taken into account. Thus, in this study, we investigate the current status of location privacy preserving in popular GLBT-focused dating applications to have a clearer view on the issue, and observe how it is being protected in the real-life practice under both already-known threat (*i.e.* trilateration) and its enhanced version (*i.e. colluding trilateration*) proposed by our group.

### B. Privacy Concerns in GLBT-focused Applications

First of all, it is important to emphasize that it is not because of hatred or discrimination that makes us opt for investigating GLBT-focused applications like Grindr, Jack'd and Hornet. But, because of their popularity, possession of highly-sensitive information, and the huge number of users[1] that make these applications highly vulnerable to cyber-attack like the case of Ashley Madison [6]. In addition, it is also because GLBT-focused dating applications like Grindr, Jack'd, and Hornet are location-sensitive, and their users depend on the publicly shown distance information to look for nearby people to meet up right away for hookup (*most of the time*), thus potentially exposed to the risk of being located.

As stated in [7], there are still many Islamic nations where homosexuality carries the death penalty. Most recently, there were several gay men in Syria lured by ISIS terrorists to go out on dates, and later executed publicly by stoning as reported in [8]. Even in those regions like North America and Western Countries, which are thought to be more open-minded, the GLBT community is still not widely accepted. More or less, people belong to the GLBT community are still facing the problem of being attacked, harassed or discriminated [9]. Such cases show that protecting privacy of the user of GLBT-focused application is a nontrivial task, and should not be neglected by the LBS providers. Because the location information together with other information such as height, weight, age, and hobby can be used to accurately disclose the targeted individuals. Later, the compromised information from those victims such as occupation, address, or frequently visiting places, daily routines and social relationships can be used to intimidate for money, or even lead to physical harassment. At this point, it is understandable why Tim Cook says: "Privacy is a matter of life and death" since he also came out as a member of the GLBT community in October 2014 [10].

### C. Organization

The rest of this paper is organized as follows. We will introduce our experimental environment in the last part of this section. In Section II, Jack'd, Grindr and Hornet are investigated in terms of location privacy. By employing our proposed colluding-trilateration, we will demonstrate how the user's location still can be accurately discovered even when countermeasures like location anonymization and location obfuscation have been implemented. In Section III, other privacy concerns are discussed with real life experiments. In this section, we will also introduce a side-channel attack fashion that can be conducted due to the current design of Jack'd. Finally, from the viewpoints of both LBS provider and user, we then give some possible solutions, and wrap up the paper in Section IV.

### D. Experiment Setup

The trilateration threat model actually can be conducted in a physical way that the adversary carries his device around to three different places and notes down the distances shown from his position to the victim. However, in order to have an easily manageable experimental environment, we employ three virtual machines that host Android OS to play the role of adversaries. Each machine is then set to be in positions around our institute as follows:

- Victim is an account run on a real iPhone 5, locates at Science Frontier Laboratory, Kyoto University with coordinates (35.02350485, 135.77687703).
- Adversary A1 is located at Demachi-yanagi Station with coordinates (35.03051251, 135.77327415).
- Adversary A2 is located at Heian Shrine with coordinates (35.01598257, 135.78242585).
- Adversary A3 is located at Kyoto Imperial Palace with coordinates (35.02258561, 135.76493382).

Each Android machine is then equipped with Fake-GPS[2] so that their positions can be freely set to any corner of the world. At the time of writing this paper, we did our experiments with Grindr (version 2.2.8), Jack'd (version 3.3.2), and Hornet (version 2.7.1). Next, to capture packets in Subsection III.A, we set up a proxy machine, and use Microsoft Network Monitor (version 3.4) to monitor network traffic passing through that proxy machine. All of the maps used in this study are sketched using a map tool available at: http://obeattie.github.io/gmaps-radius/.

## II. LOCATION PRIVACY CONCERN

To initially test whether an application adopts location obfuscation to obscure the publicly shown distance of its user or not, we move around two accounts run in the virtual environment mentioned in Subsection I.D. The distance shown on each account is then recorded and compared with the real distance. From some preliminary results, we found that Grindr and Jack'd (*in default setting*) do not adopt any location randomization, but show the exact physical distance of the user. As a consequence, the real location of the user is vulnerable to the trilateration threat model, which will be discussed in more detail in Subsection II.A. To prevent the risk of being located for its user, Grindr has already implemented a function which allows the user to hide the distance from being viewed by other users, while Jack'd has not implemented any effective countermeasure to alleviate this risk. Nonetheless, in Subsection II.B, by deploying our proposed colluding-trilateration method, we will demonstrate that disabling the "show distance" function still cannot effectively mitigate the risk. In contrast with Grindr and Jack'd, Hornet seems to be better in protecting its user's location privacy by adopting location obfuscation in its system. As a result, we always get the distance shown on the application different from the real distance. However, in Subsection II.C, we will show how our colluding-trilateration model can still be applied to precisely locate Hornet user regardless of whether location anonymization and location obfuscation are enabled at a same time in Hornet.

---

[1] According to [22], both Grindr and Jack'd currently have more than 5 million active users, while there are more than 4 million active users in Hornet. To examine that, at the time of writing this paper, we tried inputting the terms such as "gay dating" or "gay hookup" in to Appcrawlr, which is a semantic mobile application discovery powered by Softonic. The search engine indeed returned Grindr, Jack'd and Hornet on the top of the result list sorted in the order of number of downloads.

[2] https://play.google.com/store/apps/details?id=com.lexa.fakegps

*A. Trilateration Model*

As far as we are aware, the trilateration threat model (*aka: triangulation*) is said to be first reported to Grindr in 2014 [11], and discussed in recent studies [4] and [5]. The main idea of this attack model bases on the distance from the user to the adversary, which is publicly shown to other users. With privilege no more than an ordinary user, an adversary just needs to move around the victim to three different places. Distances from the victim to the adversary at those three positions are then used to pinpoint the exact location of the victim. As shown in Figure 1, Grindr and Jack'd users, who keep the default setting, are facing a high risk of being located since the adversary can obtain an accurate location up to the victim's building as highlighted in the red rectangle. With this threat model, we could not locate the victim in Hornet because the real distance is obscured, and the publicly shown distance is changed to new value every time we re-query it. Let us revisit Hornet in Subsection II.C.

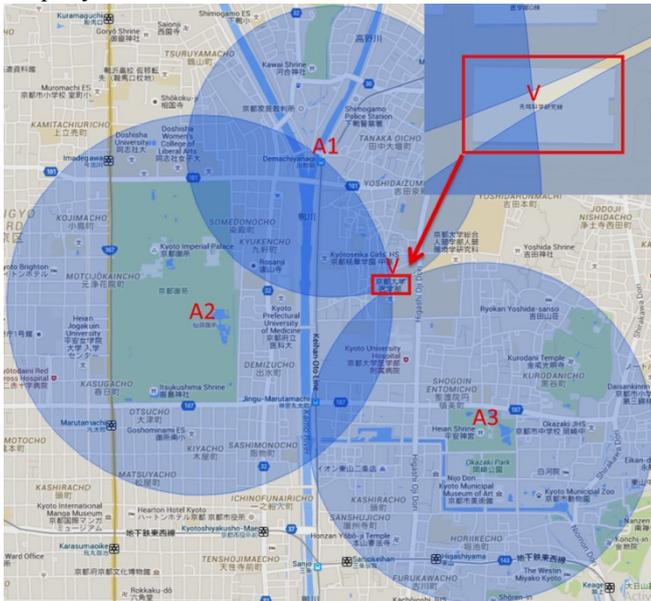

Fig. 1  Testing Trilateration Threat Model in Grindr and Jack'd.

From the geometry point of view, the location of the victim is nothing else but the coordinates of V, which is the solution (x, y) of a system of simultaneous circle equations.

$$\begin{cases} (x - x_{A1})^2 + (y - y_{A1})^2 = D1^2 \\ (x - x_{A2})^2 + (y - y_{A2})^2 = D2^2 \\ (x - x_{A3})^2 + (y - y_{A3})^2 = D3^2 \end{cases}$$

Where:
- $(x_{A1}, y_{A1})$, $(x_{A2}, y_{A2})$ and $(x_{A3}, y_{A3})$ are latitudes and longitudes of the adversary at three different positions A1, A2, A3 respectively.
- D1, D2, and D3 are distances from V to A1, A2, and A3 respectively.

In response to this type of threat, Grindr has adopted a function in which the user can opt to hide the distance since August 2014 [12]. Thus, the trilateration model is no longer able to locate those users who already disabled the "show distance" function. As we revisited [11] at the time of writing this paper, the map is no longer able to pinpoint Grindr users as shown in Figure 2.

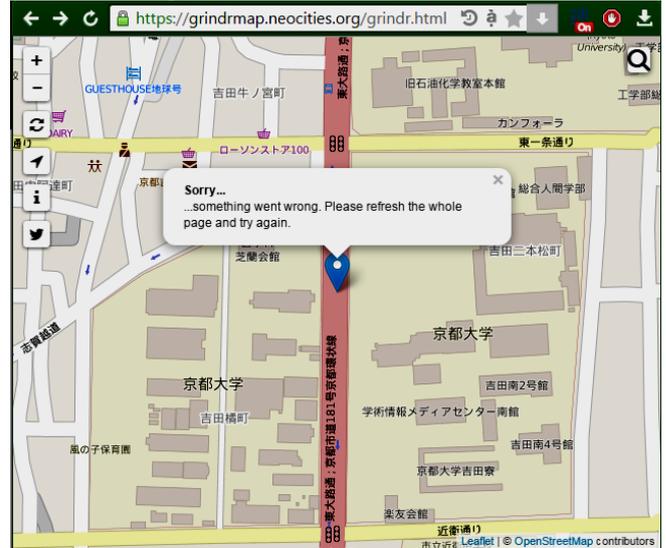

Fig. 2  Previous Grindr's flaw had been fixed.

For Jack'd, it does not adopt the location anonymization policy to protect its user. Instead, it creates a function which allows its user to adjust the accuracy of the distance to three levels: close, near, and far (*in iOS*); or street, neighborhood, and city (*in Android*) as shown in Figure 3.

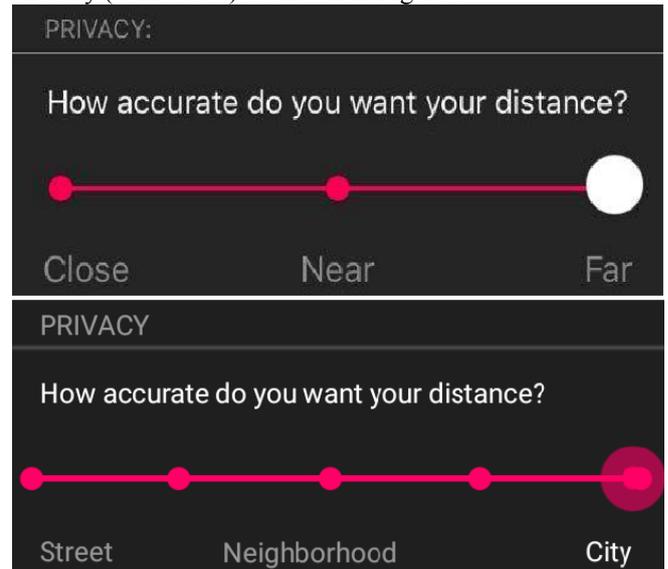

Fig. 3  Jack'd Privacy Setting.

Notwithstanding these setting options, the publicly shown distance between two of our fake accounts does not change even when we restart the application to load the new privacy setting. As a result, this new function of Jack'd does not guarantee the location privacy of its user at all.

*B. Your Neighbors are My Spies – Colluding Trilateration*

Despite of the fact that the best solution to protect the location information is not to publish it; in this part, as a key point of this paper, we will illustrate an enhanced version of the trilateration threat model that current approach like location anonymization implemented by disabling the "show distance" function still cannot effectively counter to. The primary factor in the success of this threat model bases on the way that Grindr arranges its users on the screen. Perhaps, in order to provide a high utility for the application, users are displayed left-to-right and top-to-down in an ascending order of their distances regardless of whether they have already disabled the "show distance" function or not. By exploiting this fact, the two neighbors appear just before and just behind the victim on the application's screen unintentionally become the upper and lower bounds of the distance from the victim to the adversary. As a result, the region in which the victim is locating is easily obtained by employing the trilateration model again, but with two circles drawing from the adversary to the two nearest neighbors as shown in Figure 4.

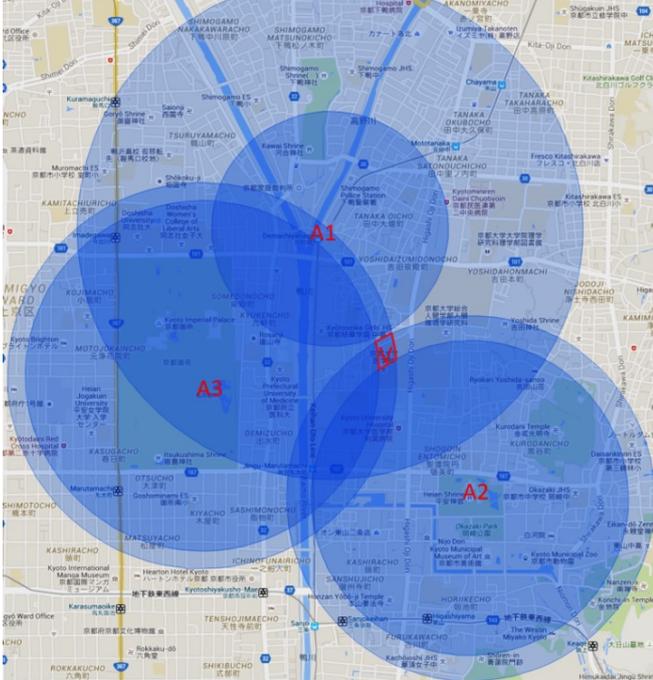

Fig. 4  Enhanced version of Trilateration Threat Model.

By using this model, the adversary even does not need to view the victim profile three times to record the distance as done in the original trilateration model, but still very effective at locating the victim's region as marked in Figure 5.

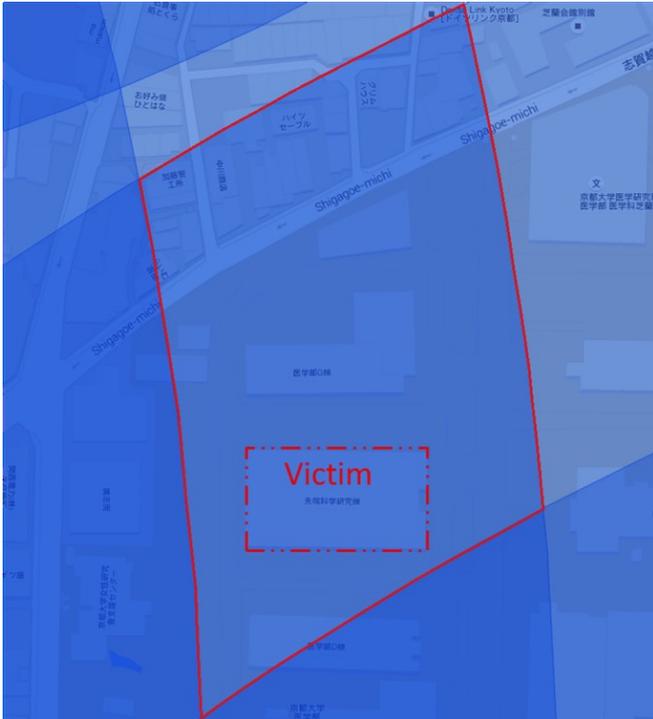

Fig. 5  The victim's region is smoothly bounded.

Hence, instead of querying the distance information of the victim several times, the adversary just needs to query it from the two nearest neighbors (*appear on the screen of Grindr*) of the victim from three different points, once for each. As a result, no distance query from the adversary to the victim is issued in this enhanced model. Therefore, approach like limiting the number of queries issued to get the distance information from one user to another user is not adequate in this case.

Moreover, a lesson learned from Figure 4 and Figure 5 is that the adversary gains more location information about the victim at A2 and A3 than A1, because the bounds set by the pairs of victim's neighbors from the viewpoints of A2 and A3 are narrower than from A1. Therefore, even without drawing the pair circles from A1, the adversary can still confidently infer the possible region of the victim, because one of the two intersection areas is a river space, thus the probability that the victim is locating in that region is relatively low. Since discussing about the probability of possible activity region of smart phone user is beyond the scope of this paper; for more information, the reader can refer to [13], in which the probability of possible locations of smart phone user has already been discussed.

So far, one may think that this type of attack model is not valid in case the victim locates in low-density area, because there are not so many neighbors around him for this attack model to take place. However, as far as we are concerned, using the term "victim's neighbors" actually is not always correct, because they may not physically locate near to the victim in real world. In fact, this attack model is still valid as long as the following condition holds:

$$AN1 < AV < AN2$$

Where:
- AV is distance from the adversary to the victim.
- AN1 and AN2 are distances from the adversary to the pair of the so-called victim's nearest neighbors.

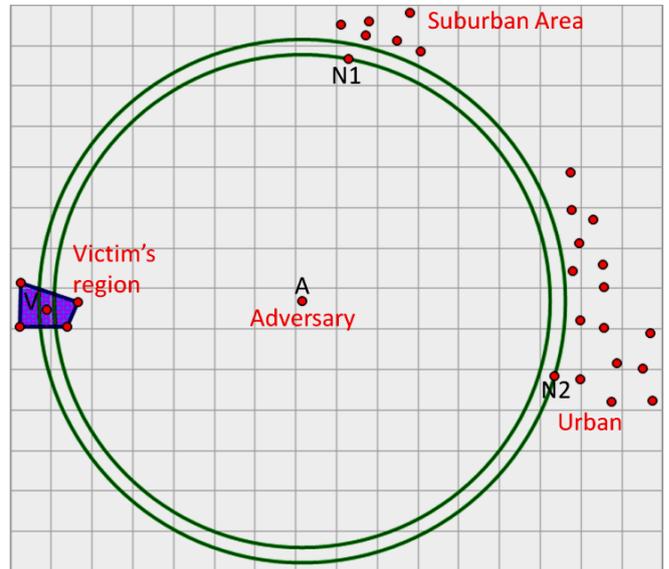

Fig. 6  Victim's neighbors are not necessarily close to him.

In real life attack, the adversary can apply this model to attack the victim in remote area by placing himself in the middle of high-density areas and the victim's region as shown in Figure 6. The more crowded the urban areas are, the more resources that the adversary can obtain to precisely explore the victim's location. Or, in a more active attack fashion, the adversary can create two colluding accounts and move them around until he can satisfactorily compromise the victim's location. The key idea is to gradually reduce the subtraction value of |AN1-AN2| such that V is still sandwiched by N1 and N2 on the application's screen of the adversary. The smaller the subtraction value becomes, the more accurate location of the victim can be revealed. Up to this point, it is obvious that obtaining victim's real location from Figure 5 becomes a trivial task. With this enhanced model, even when all local members hide their distances, the adversary can still make completely use of his colluding fake accounts to infer, thus be able to narrow down the possible region in which the victim is locating. That is where the name "colluding trilateration" of our proposed method originates from. The idea is demonstrated in Figure 7.

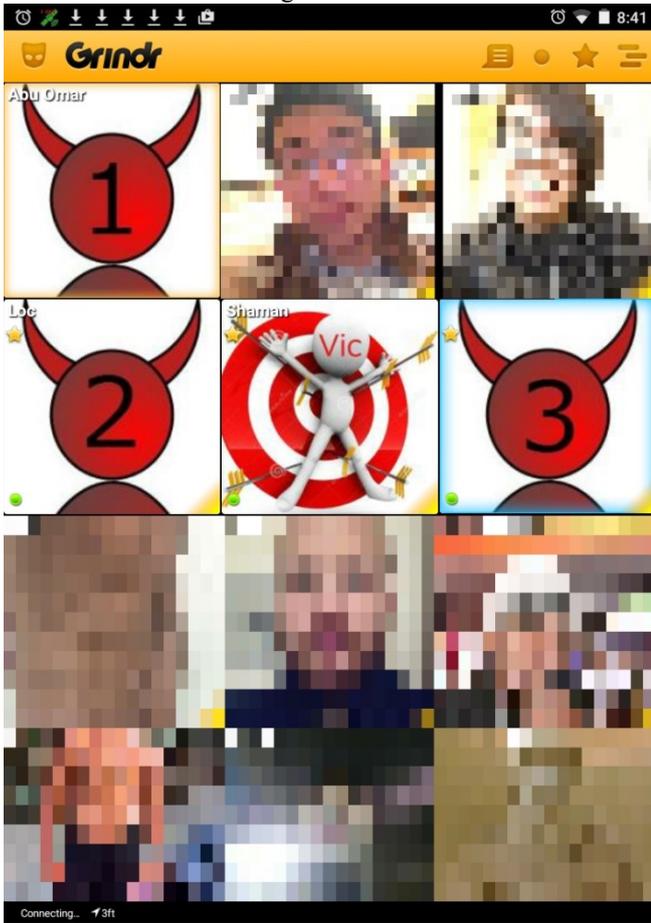

Fig. 7 Colluding Trilateration in Grindr.

As illustrated, the adversary, whose account is marked with 1, can create two colluding accounts 2 and 3, then positions 2 and 3 in a way that the victim (*Vic*) is sandwiched between 2 and 3. Later, the adversary can gradually move 2 and 3 so that the value of the distance between them becomes smaller while the victim remains in between of 2 and 3 on the screen, thus be able to precisely figure out the distance of the victim basing on the distances shown in the profile of the colluding accounts 2 and 3.

### C. Location Obfuscation is not enough

While Grindr adopts location anonymization approach to protect its users from the trilateration threat model, Hornet adopts both location anonymization and location obfuscation. In other words, all of the publicly shown distances in Hornet are obfuscated as mentioned in our preliminary result above. Even from a same location, we keep receiving different distance values every time we issue a new query to reload the profile page of the victim. Therefore, we can initially confirm that Hornet does not use a one-to-one function to obscure the real distance. In addition to this feature, Horner user can also choose to hide his obfuscated distance from other users. As a result, Hornet seems to be better in protecting its user's location from the attack fashion of trilateration model.

Indeed, Hornet was first released in 2011[3], later than Jack'd (2010)[4] and Grindr (2009)[5], thus carefully designed with the concern of privacy and security issues [14].

Because Grindr is one of the first GLBT-focused applications introduced to the society, and widely used in North America, it has become a typical object for many studies ranging from privacy, psychology, gender to sexuality health and so on; while Hornet is new and has not been thoroughly studied, especially in the aspect of user privacy and information security. For instance, Hornet was recently discussed in [15], but the authors did not accurately study the distance obfuscation pattern of Hornet. Instead, they stated that Hornet sends the distance with 10 m accuracy. However, it is not true, since Hornet does carefully obfuscate the distance of its user as stated on its homepage [16]. Therefore, to the best of our knowledge, we are the first group that empirically studies the location privacy aspect of Hornet.

In order to discover the location obfuscation pattern of Hornet, we repeatedly move around all of our adversary accounts to about 3000 different locations within 3 Km from our institute, and record both real distances estimated by ourselves and obfuscated distances shown in our Hornet accounts. Next, we scatter the data and obtain the graph in Figure 8.

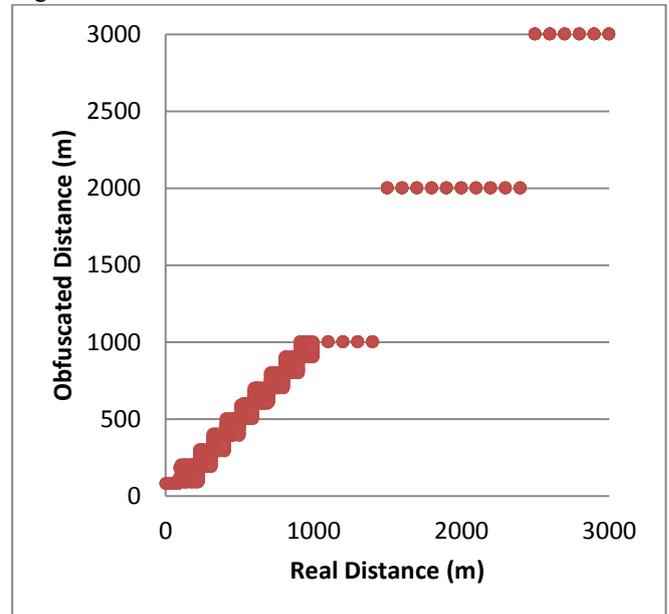

Fig. 8 Hornet Location Obfuscation Pattern.

As we already noticed that Hornet does not simply use a one-to-one function to obscure the distance, we cannot just easily employ the regression analysis to figure out the location obfuscation pattern used in Hornet. By observing the recorded data, we found that Hornet is very careful in designing its location obfuscation scheme with three different strategies for three different ranges of distance, which are 0~100 m, 100~1000 m, and above 1 km. Within the range of

---

[3] https://www.appannie.com/apps/all-stores/app/hornet/
[4] https://www.appannie.com/apps/all-stores/app/jackd/
[5] https://www.appannie.com/apps/all-stores/app/grindr/

0~100 m, Hornet makes the publicly shown distance equal to 80 m for all of the distances which are shorter than 80 m. For distance longer than 80 m and shorter than 100 m, Hornet arbitrarily adds some noise to the real distance such that the obscured distance varies within 80 m and 100 m. We break down the Figure 8 to have a clearer view on the distance obfuscation pattern in the range of 0~100 m.

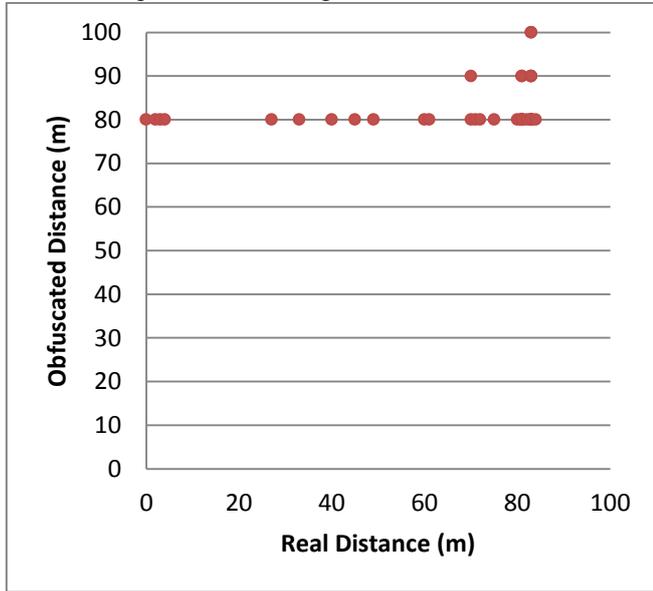

Fig. 9 Hornet Location Obfuscation Pattern – Range 0~100m.

Next, within the range of 100~1000 m, Hornet adopts a more complex obfuscation pattern that changes the publicly shown distance every time we refresh the victim profile to get the distance information. As a result, although the positions of our fake accounts did not change, we always got different values shown in our fake accounts. To deeply analyze this obfuscation pattern, we issue 30 queries from every single position and note down all the possible obfuscated results returned by Hornet. Breaking down the Figure 8 in the range of 100~1000 m gives us a clearer view.

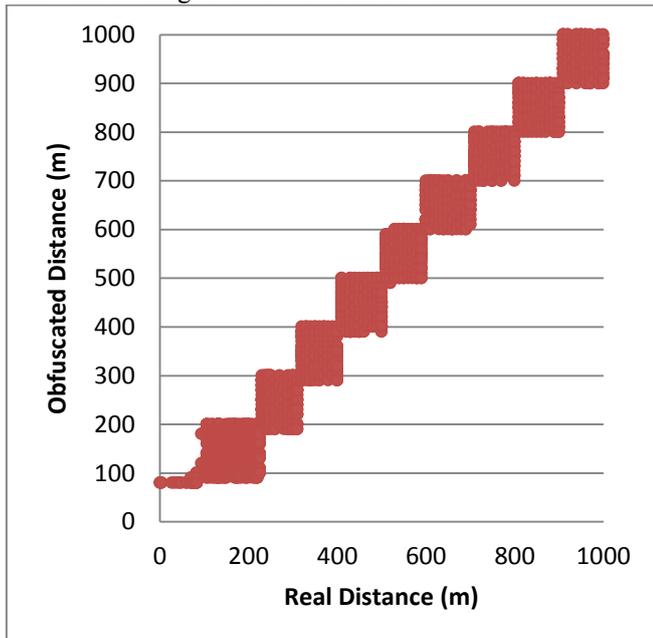

Fig. 10 Hornet Location Obfuscation Pattern – Range 100~1000 m.

From Figure 10, we can conclude that in the range of 100~1000 m, Hornet evenly randomizes the real distance with amplitude of 100 m. In more detail, a real distance, which is longer than 100 m and shorter than 1000 m, is first rounded to the nearest hundred. It is then obfuscated by adding an arbitrary number ranging from 0 to 100 with the step of 10. For example, if the real distance is 321 m, then the obfuscated distance is 300 plus any random value in the set of {0, 10, 20, 30, 40, 50, 60, 70, 80, 90, 100}. Thus, the publicly shown value can be any number between 300 m and 400 m with the step of 10.

Finally, for distance longer than 1 km, Hornet applies another obfuscation pattern in which the real distance is rounded to the nearest one in the unit of km. For instance, 1.2 km is rounded to 1 km, while 1.6 km is rounded to 2 km.

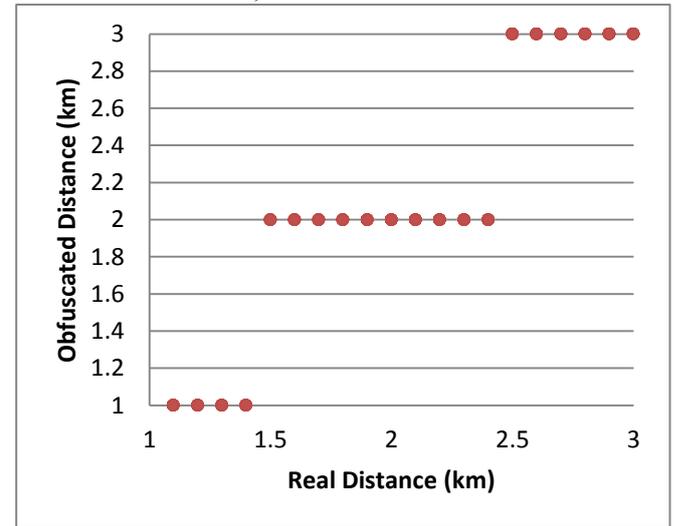

Fig. 11 Hornet Location Obfuscation Pattern – Range 1 km ~.

From the patterns discussed above, it is not an exaggeration to say that the obfuscation used in Hornet is very reasonable with respect to the privacy preserving for Hornet user. Distance of every user in the range of 0~1000 m is complexly obfuscated because this range is considered to be very sensitive to the user's location privacy; while distance longer than 1 km is simply obscured by rounding method because it is far enough to protect the user privacy. Also, showing the approximated distance for those users, whose distance is longer 1 km, does not have any bad impact on the utility of other local users. Therefore, the obfuscation does not only provide Hornet user with a better protection, but also makes it more difficult for the adversary to carry out the trilateration attack. That is the reason why we could not find out the exact location of the victim in Hornet when applying the trilateration model because three circles sketched from three different locations of our adversary accounts do not converge on one point.

Nevertheless, we still have the colluding-trilateration method proposed in Subsection II.B to test in Hornet; because, same as Grindr, Hornet also arranges the users on the screen from left-to-right and top-to-down in an ascending order of their distances regardless of whether they have disabled the "show distance" function, and the publicly shown distance has already been obfuscated. Therefore, we reuse the colluding-trilateration to examine whether we can discover the real location of the victim or not. This time, we just move accounts 2 and 3 to obtain a same result as shown in Figure 7, but do not use the distances shown on these two fake accounts because they are already obfuscated by Hornet, thus not correct. However, because those fake accounts belong to us, we can always measure the real distances between them as side-channel knowledge without relying on the one shown in Hornet.

Surprisingly, Hornet perhaps might have already envisioned about this type of attack and took a step ahead. What Hornet does is to randomly remove some users from the screen of other local users. In other words, at some positions, on the screen of account 1, we can see all of the victim (*Vic*) and accounts 2 and 3; but at some other positions we cannot. That means Hornet does not always show all surrounding local users on one's screen. Instead, it arbitrarily drops some users. As a consequence, we cannot set the upper and lower bounds for the victim's distance to conduct the colluding-trilateration.

In order to bypass this countermeasure, we make completely use of the Favorites List of Hornet. Similar to most of other Social Network platforms, Hornet also has a feature in which a user can add other users to his favorite list like the "follow" function in Twitter. Then, there is a separate tab for the user to view his Favorites List in which the favorite users in this list are also arranged from left-to-right and top-to-down in an ascending order of their distances regardless of whether they have disabled the "show distance" function, the publicly shown distance are obfuscated, or they are not shown on the local screen of other users. The trick is demonstrated in Figure 12. (*Two accounts at the top-left corner are added by default for all users in Hornet to promote sexuality health.*)

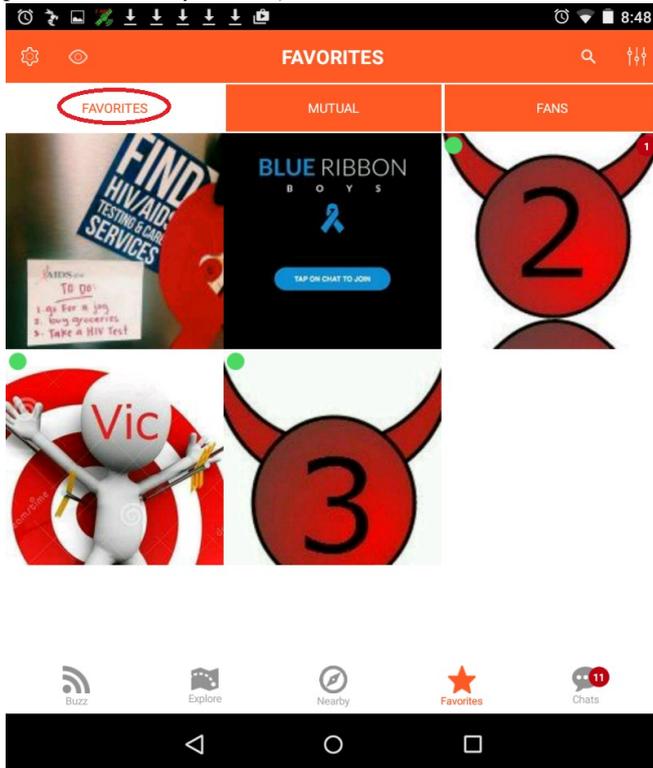

Fig. 12 Colluding Trilateration in Hornet bases on Favorites List.

Taking advantage of this Favorites List, the adversary first needs to add the other two fake accounts 2 and 3 to his Favorites List, and then also adds the targeted victim into this list at the first time he sees the targeted victim on the public local screen. That is how the adversary can anchor the victim even when the victim does not appear on his local screen. Finally, the locating problem becomes a same problem as shown in Figure 4. As a result, we were able to locate the position of the victim (*Vic*) in Hornet although all the distances are obfuscated and the victim already disabled the "show distance" function in his application.

## III. OTHER PRIVACY CONCERNS

### A. Vulnerabilities from Third-Party Advertisement Banner and Misconduct in Handling User's Personal Information

In the age of online marketing, one's privacy is often threatened by the very advertisements popping up in his device as mentioned in [17] and [18]. In order to investigate this issue in Jack'd and Grindr, we analyze packets captured while using these two applications. The experiment results are shown in Figure 13.

```
229..x-failurl: http://ads.mopub.com/m/ad?v=8&udi
d=ifa:E121C9CB-9758-4076-9DE1-422BB33F3F84&id=agl
tb3B1Yi1pbmNyDQsSBFNpdGUY7cz7Bgw&nv=1.17.2.0&q=m_
gender:m,m_age:0&o=p&sc=2.0&z=+0900&mr=1&ct=2&av=
2.2.4&cn=KDDI&iso=jp&mnc=50&mcc=440&dn=iPhone5%2C
2&exclude=46a0e29f574448038760b6e06a3232f4&reques
t_id=e7918915a00144508c4afd5fa7fcbb91&fail=1..x-i
mptracker: http://ads.mopub.com/m/imp?appid=&cid=
31fd6d9c46d14787b072ca69dee8b44c&city=&ckv=2&coun
try_code=JP&cppck=2375D&dev=iPhone5%2C2&id=agltb3
B1Yi1pbmNyDQsSBFNpdGUY7cz7Bgw&is_mraid=1&mpx_clk=
http%3A%2F%2Fmpx.mopub.com%2Fclick%3Fad_domain%3D
agoda.com%26adgroup_id%3D46a0e29f574448038760b6e0
6a3232f4%26adunit_id%3DagltB3B1Yi1pbmNyDQsSBFNpdG
UY7cz7Bgw%26ads_creative_id%3D31fd6d9c46d14787b07
2ca69dee8b44c%26app_id%3DagltB3B1Yi1pbmNyCwsSA0Fw
cBiJwSEM%26app_name%3DGrindr%2520iOS%26auction_ti
me%3D1444444014%26bid_price%3D19.49%26bidder_id%3
4..x-failurl: http://ads.mopub.com/m/ad?v=8&udid=
ifa:BBC656C1-3F0B-4B6A-8D85-77D7E1D5476C&id=d7ea3
f8c3825497f940bf56b05335665&nv=3.3.0&o=p&sc=2.0&z
=+0900&ll=35.02353627550613,135.776885205088&lla=
65&llsdk=1&mr=1&ct=2&av=3.1&cn=KDDI&iso=jp&mnc=50
&mcc=440&dn=iPhone5%2C2&ts=1&request_id=cdacc5b2d
47445098c877eb5b4202bd1&fail=1&fail=1&exclude=628
0a0eef1a14dd58f421e5c4cc0a94b&exclude=74de412afe2
611e38aab1231392559e4&exclude=75092462fe2611e38aa
b1231392559e4&fail=1..x-height: 50..x-imptracker:
 http://ads.mopub.com/m/imp?appid=&cid=03faca92f1
4c4f6b9d6896069663eb18&city=&ckv=2&country_code=J
P&cppck=FC310&dev=iPhone5%2C2&id=d7ea3f8c3825497f
940bf56b05335665&is_mraid=0&mpx_clk=http%3A%2F%2F
mpx.mopub.com%2Fclick%3Flineitem%3D5217644%26ad_d
omain%3Dwish.com%26ad_id%3D666eba282b3623b1%26adg
roup_id%3D6280a0eef1a14dd58f421e5c4cc0a94b%26adun
it_id%3Dd7ea3f8c3825497f940bf56b05335665%26ads_cr
eative_id%3D03faca92f14c4f6b9d6896069663eb18%26ap
p_id%3D09f13c886c604d25a5245842158819B9%26app_nam
e%3DJack%25E2%2580%2599d%2520-%2520iOS%26auction_
time%3D1444398521%26bid_price%3D0.13%26bidder_id%
```
Fig. 13 : Information Leak through Third-Party Ads.

Surprisingly, in both applications, the third-party advertisements leak many important information of the user including name of Telecommunications Service Provider (i.e. KDDI), device's model information (i.e. iPhone 5), country code (i.e. JP), and last but not least: the name of the applications (i.e. Grindr and Jack'd) which is the most sensitive information that no any straight-acting person wants the others to know the existence of such applications in his phone. It may have no problem if the packets are sent directly to the ads provider's server. However, what is worth mentioning here is that the packets are sent in an unencrypted fashion, thus widely open to an attack type known as man-in-the-middle attack, in which the hacker taps the Internet connection of the victim to eavesdrop the packets.

By analyzing all the captured packets, we were further shocked by the fact that all the packets containing members' profile pictures of Grindr are also sent in the air without encryption, thus being captured and recovered back to the original image files as shown in Figure 14.

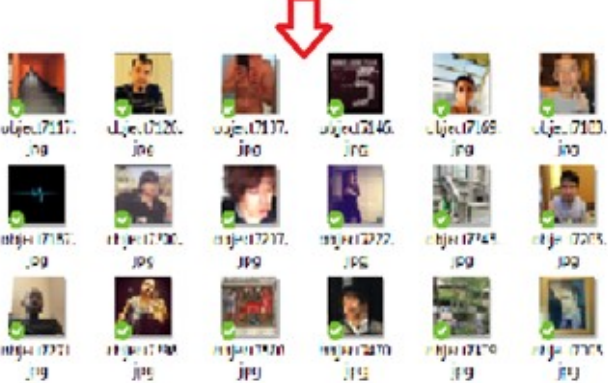

Fig. 14 Pictures recovered from unencrypted packets in Grindr.

Concerning ethical issue, we only recover image files that appear on the first screen page of Grindr as evidence, and the users' avatars are intentionally censored. For Jack'd, it is very careless in handling its user's private photo, because even when a photo is sent in a private message, Jack'd does not use any secure connection to protect the photo. Instead, Jack'd sends it via HTTP, which is an unsecure transmission protocol as shown in Figure 15.

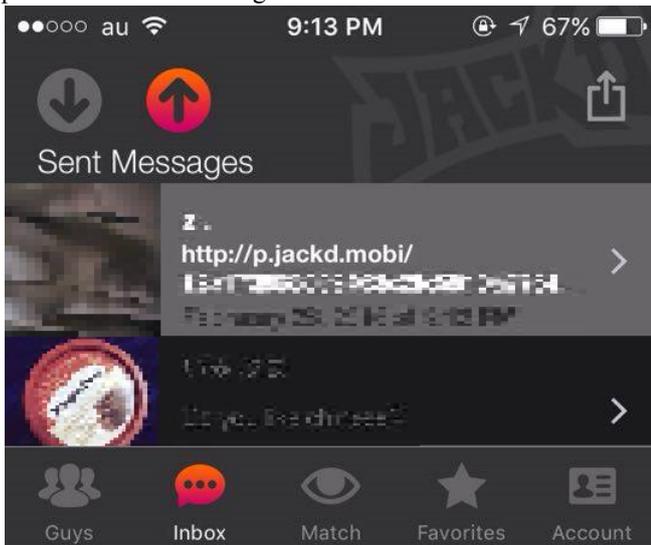

Fig. 15 Unsecure http protocol used in Jack'd to send private photos.

Taking into account of the above findings, it is obvious that the user's privacy is not guaranteed at all although the vendor has been alerted to these issues by a security firm before [11].

As for Hornet, it is not necessary to do this experiment, because it has been already confirmed in [14] and [15] that Hornet carefully employs SSL certificates and HTTPS protocol for its connection.

*B. Together with IP Spy and Linkage Attack*

Next, as human being is born curious, social engineering intrusion techniques like phishing is always the easiest but effective way to compromise people. Since Jack'd provides it users with a feature to see who viewed his profile with timestamp, an adversary can put an IP-spy URL into his profile to promote his appearance, thus being able to obtain the victim's IP address if the victim feels curious and clicks on that URL. Nevertheless, as it was discussed in [19] that IP address is also important personal information which can be exploited to perform the linkage attack to retrieve other personal information. To have a clearer view on how this gimmick is really effective at luring innocent users, we place our Jack'd accounts in three big cities of Japan which are Tokyo, Osaka and Kyoto within 12 hours (*from 6PM to 6AM of the following day*) to estimate how many innocent and curious victims could be lured. We choose this time period because it is the most active usage time according to [20]. To conduct this task, we had to reboot the virtual machines every two hours so that our accounts will not disappear from the screen of other local users, as we found that Jack'd only keeps an account displayed on other user's screen for two hours since the latest login. The result is illustrated in Figure 16.

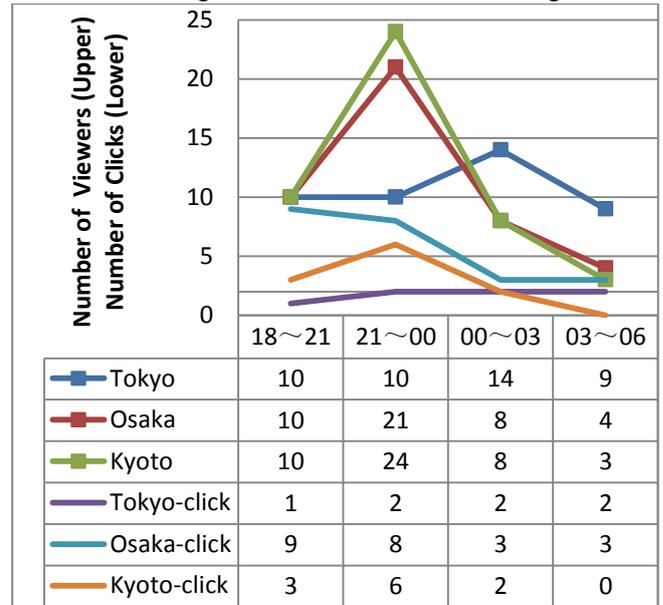

| | 18〜21 | 21〜00 | 00〜03 | 03〜06 |
|---|---|---|---|---|
| Tokyo | 10 | 10 | 14 | 9 |
| Osaka | 10 | 21 | 8 | 4 |
| Kyoto | 10 | 24 | 8 | 3 |
| Tokyo-click | 1 | 2 | 2 | 2 |
| Osaka-click | 9 | 8 | 3 | 3 |
| Kyoto-click | 3 | 6 | 2 | 0 |

Fig. 16 Analysis from IP-spy Intrusion Gimmick.

In total, we got 131 viewers from three accounts with 41 viewers clicked through the IP-spy URL we put in the profile.

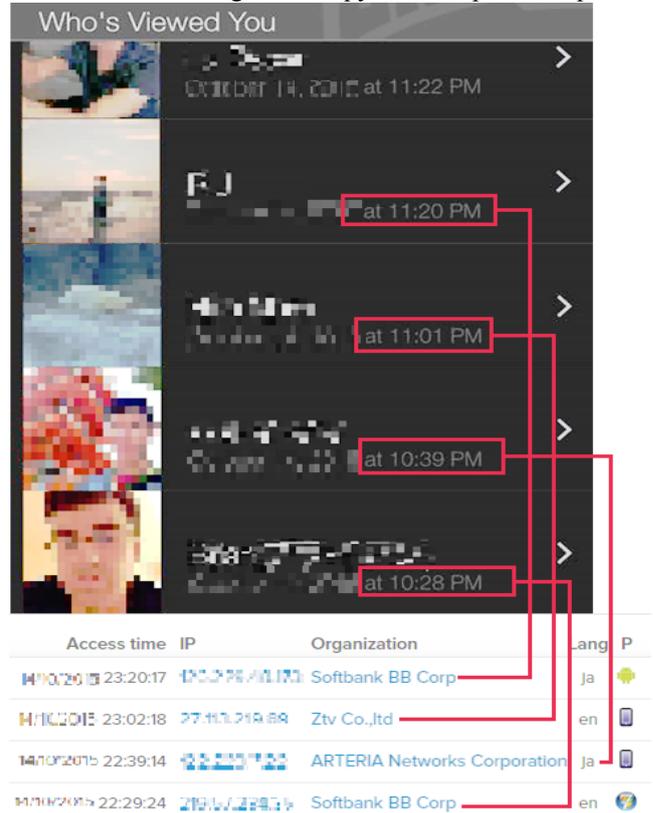

Fig. 17 Linkage Attack with IP-spy and Jack'd timestamp.

Among these 41 clicks, we were able to perform linkage attack to 26 users with high confidence by matching the timestamp between Jack'd and our IP-spy server to further reveal other information including their IP address, ISP, display language and platform of their devices as shown in Figure 17.

IV. DISCUSSION AND CONCLUSION

Through this study, we would like to particularly alert the users of Grindr, Jack'd, and Hornet as well as the users of other LBS in general about the risk of being located easily regardless of whether the recent location anonymization and location obfuscation approaches have been adopted. By investigating these three applications, we found a paradox that although there have been many attack models proposed by the privacy-preserving researchers, the user's location privacy has not been seriously taken in to consideration by the LBS provider and the user themselves. As far as we are concerned, the reason of this negligence derives from both sides. From the viewpoint of the LBS provider, it might cause overhead to implement those sophisticated solutions proposed by the research community, while the utility of the application is not really guaranteed, thus probably lead to the loss of its customer. From the viewpoint of the user, it is maybe because of two reasons. Perhaps, the first one is also due to the trade-off of utility. The other one is because of unawareness. For that reason, in this paper, instead of using complicated mathematical equations and complex algorithms to show the threat models, we opt for visualizing it on maps and figures so that even those non-technical readers can understand how easily their privacy can be compromised in the current security condition.

In order to alleviate the risks of man-in-the-middle attack, IP spy and other side channel attacks as mentioned in Section III, we urge the LBS providers and involved third-parties to carefully encrypt the connection from their servers to the users. For the user, we suggest not opening any URL out of curiosity. If it is really necessary to open an unknown URL, the user should turn on VPN at first to prevent the leak of their real IP address.

For those threat models discussed in Section II, let us argue that privacy preserving policy is different from person to person. Especially in GLBT community, some already came out, thus have no concern about privacy; while some are straight-acting, thus do not want to be disclosed. Therefore, a centralized solution is not really suitable, and users are the very ones who need to make decision whether to protect their own privacy. For the meantime, while waiting for the experts and the LBS providers to discover a perfect solution for location privacy protection without trading off the utility, we suggest that the user should take a step ahead to protect their own privacy from those vulnerabilities mentioned in this study. That is to use Fake-GPS applications like the one that we use in this study (*probably also used by most of the adversaries*) to hide the real location to an acceptable extent so that the user can still gain the convenience provided by the LBS. How far the fake location should be shifted from the real one depends on how much utility and convenience that a user is willing to trade off with his privacy, thus different from case by case. We strongly believe that this user-centric solution not only suits all type of users, but also helps to save the vendors from overhead investment in implementing sophisticated solutions and infrastructures.

Apart from technical methods aforementioned, human factor is also important in protecting oneself in the cyberspace. In order to avoid troublesome problems in the future when the vendors get hacked as the case of Ashley Madison [6], the user should not register account to those highly sensitive applications under his real name or even a part of his real name. Instead, the user should use information that could not be used to link the account with his real-life personally identifiable information.

Last but not least, in this paper, we of course could have utilized more complicated techniques to extract and test the accuracy of the threat models with more users of Grindr, Jack'd, and Hornet in bulk. However, as far as we are concerned with the ethical issue that those compromised users also have their right to be undisclosed, and there may be our acquaintances among them. We thus did not go beyond those accounts created by ourselves.

ACKNOWLEDGMENT

This work was supported by JSPS KAKENHI Grant Number 15K00423 and the Kayamori Foundation of Informational Science Advancement.

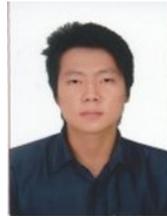

**Nguyen Phong HOANG** was born in Tien Giang Province, Vietnam in 1992. He received his undergraduate degree in Business Administration majoring in Information & Communications technology (ICT) from Ritsumeikan Asia Pacific University, Japan. He is presently pursuing his graduate studies at the Graduate School of Informatics, Kyoto University, Japan. His research interests include information security, privacy and anonymous communication. He hopes to advance his research on Tor (The Onion Router), one of the most robust anonymous tools, during his graduate studies. He has participated in annual IEEE International Conference on Advanced Communication Technology (ICACT) since 2014. He also received the Outstanding Paper Award from the Technical Program Committee of the conference in the 16[th] and 18[th] ICACT. He has been an IEEE member since 2013, and DBSJ since 2014.

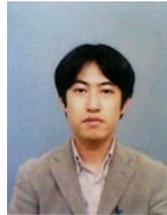

**Yasuhito ASANO** received the BS, MS, and DS degrees in information science from the University of Tokyo in 1998, 2000, and 2003 respectively. In 2003-2005, he was a research associate in the Graduate School of Information Sciences, Tohoku University. In 2006-2007, he was an assistant professor in the Department of Information Sciences, Tokyo Denki University. He joined Kyoto University in 2008. He currently serves as an associate professor in the Graduate School of Informatics. His research interests include web mining, network algorithms. He is a member of the IEICE, IPSJ, DBSJ, and OR Soc. Japan.

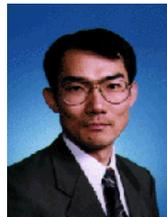

**Masatoshi YOSHIKAWA** received the BE, ME, and PhD degrees from the Department of Information Science, Kyoto University in 1980, 1982, and 1985, respectively. From 1985 to 1993, he was with Kyoto Sangyo University. In 1993, he joined the Nara Institute of Science and Technology as an associate professor in the Graduate School of Information Science. From April 1996 to January 1997, he was in the Department of Computer Science, University of Waterloo as a visiting associate professor. From June 2002 to March 2006, he served as a professor at Nagoya University. From April 2006, he has been a professor at Kyoto University. His current research interests encompass database technologies and their application to medical healthcare domains. He is a member of the ACM, IEICE, IPSJ and DBSJ.